# The use of Lorentz microscopy for the determination of magnetic reversal mechanism of exchange-biased $Co_{30}Fe_{70}$/NiMn bilayer

A. Masseboeuf [†,*], C. Gatel*, P. Bayle-Guillemaud

*CEA-Grenoble, INAC/SP2M/LEMMA, 17 rue des Martyrs, F-38054 Grenoble, France*

Y. Lamy, B. Viala

*CEA-Grenoble, DRT/LETI, 17 rue des Martyrs, F-38054 Grenoble, France*

**Abstract**

Lorentz Transmission Electron Microscopy (LTEM) combined with *in situ* magnetizing experiments is a powerful tool for the investigation of the magnetization reversal process at the micron scale. We have implemented this tool on a conventional Transmission Electron Microscope (TEM) to study the exchange anisotropy of a polycrystalline $Co_{35}Fe_{65}$/NiMn bilayer. Semi-quantitative maps of the magnetic induction were obtained at different field values by the Differential Phase Contrast (DPC) technique adapted for a TEM (SIDPC). The hysteresis loop of the bilayer has been calculated from the relative intensity of magnetic maps. The curve shows the appearance of an exchange bias field reveals with two distinct reversal modes of the magnetization: the first path corresponds to a reversal by wall propagation when the applied field is parallel to the anisotropy direction whereas the second is a reversal by coherent rotation of magnetic moments when the field is applied antiparallel to unidirectional anisotropy direction.

[†] corresponding author: aurelien.masseboeuf@cemes.fr

* New ad. : CEMES-CNRS - 29, rue Jeanne Marvig  31055 Toulouse Cedex

**Introduction**

The observation with a nanometer spatial resolution of the magnetic configurations is nowadays a great interest for many engineering applications of magnetic materials. Lorentz Transmission Electron Microscopy (LTEM) is one of the techniques which enable analysis of local magnetic properties. This technique allows *in situ* observations of the domain structure of a magnetic material at different magnetic field values [1]. Classical LTEM relies on the fact that an electron beam passing through an area with a component in magnetic induction perpendicular to its trajectory will be deflected by the Lorentz force. The magnetic induction arises either from the magnetization in the sample itself or the nonzero divergence of the magnetization which leads to stray fields exterior to the sample. In the TEM the deflection of the electron beam results from the perpendicular component of magnetic induction averaged on the electron path (sample plus vacuum).

Two methods can be distinguished to image the magnetic domain configuration in LTEM: the Fresnel mode and the Foucault mode. The Fresnel contrast appears when the Lorentz lens is defocused: the image intensity increases at the position of some domain walls and decreases at the position of others. In Foucault mode, a contrast aperture is introduced in the back focal plane of the Lorentz lens and positioned in order to intercept electrons which have passed through one set of domains magnetized in a given direction. The contrast corresponds to a dark field image where only these set of domains appear dark. An extension of this technique is the acquisition of series of Foucault images in order to get magnetic map. This technique so-called Differential Phase Contrast (DPC) was initially developed [2] on a Scanning Transmission Electron Microscope (STEM) and latter adapted on a conventional Transmission Electron Microscope (TEM) [3]. This latter technique so-called SIDPC (SI for Series of Images) consists to record series of images by moving the aperture or tilting the incident beam in a +/-X and a perpendicular +/-Y directions. It has been shown that adding the images in each direction produces two images linearly proportional to $B_X$ and $B_Y$ [3]. The magnetic components are then computed relative to the origin of the incident beam (no magnetic deflection) and the vector map can be easily displayed from the two images. Magnetic imaging for remanent state is performed with the specimen in a field free region in the TEM column by turning off the main objective lens and using another lens (so called Lorentz lens) placed below the sample as imaging lens. In order to perform *in situ* magnetization for reversal process analysis, the in-plane field applied on the sample can be created by tilting the specimen holder in the axial magnetic field of the objective lens (either residual field or field obtained by switching on the objective lens). These fields have been calibrated with a Hall effect sensor. Major advances in the comprehension of magnetic/antiferromagnetic coupling response to an external field have thus been achieved using LTEM : temperature dependence of the exchange bias field [4-5], implementation in spin valves [5], microstrucure of the ferromagnetic layer [6-7], domains formation and domain walls propagation [4,8] or growth conditions [8-9].

The aim of this letter is to demonstrate the application of SIDPC for investigations at different magnetic field values of a system, following the complete hysteresis cycle of the film. More precisely we use a technique to reconstruct the fully hysteresis loop from induction maps, giving access to magnetic properties as it was previously described by A.C. Daykin in [6]. We focused here on the exchange anisotropy coupling between a ferromagnetic (F) layer and an antiferromagnetic (AF) layer and at the same time on the mechanism of magnetization reversal. The direct exchange coupling in AF/F bilayer has attracted great interest due to its importance in the spintronic devices as spin-valves [10-12] or magnetic tunnel junctions [13]. This phenomenon discovered by Meiklejohn and Bean over 50 years ago [14,15] creates a bias field ($H_B$) corresponding to a shift of the hysteresis loop of the F layer and an increase of its coercive field ($H_C$). This behaviour is enhanced when the AF/F bilayer has been cooled under the application of a magnetic field through the ordering temperature of the AF so-called Néel temperature [16].

**Material and methods**

The system studied in this work is a polycrystalline $Co_{35}Fe_{65}$ (70 nm)/NiMn (50 nm) bilayer. NiMn is widely used as AF layer for its high crystalline anisotropy field and high blocking temperature [17,18]. $Co_xFe_{1-x}$ is a promising candidate for future recording media [19] and RF applications [20-22] due to its high saturation magnetization. For the LTEM observation, the bilayer has been directly deposited by DC sputtering on carbon coated thin film. Growth conditions and post-annealing treatment are described elsewhere [20,23].

The LTEM experiments were performed on a conventional TEM JEOL 3010 working at 300 kV with a $LaB_6$ gun and equipped with a Gatan Imaging Filter (GIF). To reach the field free conditions, the microscope was operated in Low Mag mode using the objective minilens as Lorentz lens, the main objective lens being switched off. The selective area aperture acts as contrast aperture. In order to place this aperture in a real back focal plane to get good Foucault images each intermediate lens has been set in free lens mode [24]. An acquisition script is used to drive the tilt series across the aperture [3] and to record 4 series of 20 Foucault images. The 512x512 pixels size images are acquired with a CCD camera through a Gatan Image Filter (GIF) operated in zero loss mode with a 10 eV energy selecting slit. In this case, the signal-to-noise ratio is improved compared to unfiltered images by removing the inelastic scattered electrons [25], and the magnification is increased by the GIF lenses. Due to long exposure times (10 minutes), sample drift may introduce artefacts. We have then developed a software using Gatan Digital Micrograph to correct the drift on the series of images before to computed $B_X$ and $B_Y$ components of the magnetization. Finally the 2D integrated induction map is processed.

The in-situ magnetization can be done by many ways as using a specialized magnetization sample holder [6,26]. In our case, we have used the residual vertical magnetic field of the objective lens (measured to be of 305 Oe by a Hall probe). The specimen holder was tilted up to $|\alpha_{max}| \leq 21.5°$ in order to apply a controlled in plane magnetization between ±110 Oe on the sample and so to achieve magnetic reversal processes. The 0.1° angle uncertainty of the stage leads to an accuracy of the applied field of 0.5 Oe. In this method, a substantial out-of-plane magnetic field is applied to the sample but its effect on the magnetization is highly reduced by the shape anisotropy and can be neglected in our experiment.

**Results**

Figure 1 presents a set of integrated magnetic induction maps of the $Co_{35}Fe_{65}$/NiMn bilayer obtained by SIDPC at different field values. Colour wheel indicates direction and intensity of the magnetic vector at each pixel of the image (for a better understanding we have superimposed small white vectors). The studied uniform area (10 µm x 10 µm) has been chosen far from the edges of the sample in order to avoid magnetic artefacts and to get a uniform thickness.. The magnetic field was applied parallel or antiparallel to the pinning direction defined during the cooling of the as-grown sample [10]. This easy axis is indicated by dashed lines on the magnetic map (A) of the Figure 1. 27 maps have been recorded for applied field between ±110 Oe. From the whole set of reconstructions, two hysteresis loops presented on Figure 2 have been calculated. The Y axis corresponds to the projection of all the magnetic vectors of each map along either the anisotropy direction (filled circles) or the direction perpendicular to this anisotropy direction (open circles). The treatment of each map obtained at a given field will produce a couple of points (one for each curve). This technique allows then to measure locally the hysteresis behaviour during a magnetisation process. In our particular case, the curve computed along the anisotropy direction presents the classical shape of a hysteresis loop of an AF/F bilayer with an exchange coupling. This square loop is shifted by a bias field ($H_B$) of about 30 Oe which is in a very good agreement with the ex-situ macroscopic measurement obtained on the same sample [17]. The main feature which could be extracted from these curves is the appearance of two distinct reversal modes of the magnetization for increasing and decreasing field. The first path (noted (1)) on the Figure 2 corresponds to a reversal by wall propagation when the applied field is parallel to the anisotropy direction. Different steps of this reversal mode are presented on images A, B and C on Figure 1. At first the F layer is saturated (A), then contrasts corresponding to ripple contrasts in Fresnel images become visible due to the high crystalline anisotropy of the polycrystalline $Co_{35}Fe_{65}$ layer (B) when decreasing the magnetic field. Many cross-tie walls at 180° (surrounded by dashed white circles) appear suddenly with a very fast propagation (C). The second path (noted (2) is a reversal by coherent rotation of magnetic moments when the applied field is antiparallel to the unidirectional anisotropy direction. Images D, E and F on Figure 1 reveal the process: the saturated magnetization (D) rotates coherently (E) when increasing the magnetic field until the creation of a 90° wall (F) which finally propagates to complete the reversal process. The difference between the two reversal processes is highlighted on the open circles curve which represents the magnetization along the axis perpendicular to the anisotropy direction. We can indeed observe on the right hand side of the curve a non zero value showing that magnetic moments lie perpendicular to the anisotropy axis at point F (path 2), showing the rotation process. The presence of two distinct reversal processes is explained by the unidirectional anisotropy which favours a fast reversal with wall propagation in one direction [26]. Note that in the other direction, the coherent rotation is not complete as the 90° angle between magnetic moments is energetically unfavourable compared to 180° configuration.

**Conclusion**

To conclude, we have demonstrated that LTEM is a powerful tool for investigating locally and dynamically the magnetic properties of magnetic systems. We have used the SIDPC technique with *in situ* magnetizing experiments to obtain magnetic maps at different applied field values. Hysteresis loop along any directions in the observation plane can be reconstructed from a series of maps and give quantitative values on the magnetic properties as coercive field or bias field. The advantage of the technique is to couple the measurement and the image showing the magnetic configuration: distribution of the induction vector, or presence and type of the domain walls. We have applied this method to study the exchange coupling of a polycrystalline $Co_{35}Fe_{65}$/NiMn bilayer and measured a bias field of 30 Oe. Two distinct reversal processes explained by the unidirectional anisotropy due to the exchange coupling have been analysed. More over, it is important to note that conventional TEM in free lens mode can be used easily to make such accurate observations. Furthermore, the resolution will be increased in a TEM equipped with a dedicated Lorentz lens and even aberration-corrected Lorentz lens [27]. The magnetic spatial resolution of the maps obtained on our system has been estimated around 30 nm and could reach less than 5 nm with a dedicated Lorentz lens.

# References

[1] J.N. Chapman, J. Phys. D **17**, 623 (1984)

[2] J.N. Chapman, R. Ploessl and D.M. Donnet, Ultramicroscopy **47**, 331 (1992)

[3] A.C. Daykin and A.K. Petford-Long, Ultramicroscopy **58**, 365 (1995)

[4] P. Gogol, J.N. Chapman, M.F. Gillies and F.W.M. Vanhelmont, J. Appl. Phys. **92**, 1458 (2002)

[5] X. Portier, A. K. Petford-Long, T.C Anthony and J.A, Brug, App. Phys. Lett., **75**, 1290 (1999)

[6] A. C. Daykin, J.P. Jakubovics and A. K. Petford-Long, J. Appl. Phys. **82**, 2447 (1997)

[7] J.P. King, J.N. Chapman, M.F. Gillies and J.C.S. Kools, J. Phys. D, **34**, 528 (2001)

[8] B. Ramadurai and D.J. Smith, IEEE Trans, Magn., **39**, 2732 (2003)

[9] M. N. Baibich, J. M. Broto, A. Fert, F. Nguyen Van Dau, F. Petroff, P. Etienne, G. Creuzet, A. Friederich, and J. Chazelas, Phys. Rev. Lett. **61**, 2472 (1988).

[10] B. Dieny, J. Magn. Magn. Mater. **136**, 335 (1994)

[11] W.C. Cain, W.H. Meiklejoh, and M.H. Kryder, J. Appl. Phys. **61**, 4170 (1987)

[12] J.S. Moodera, L.R. Kinder, T.M. Wong, and R. Meservey, Phys. Rev. Lett. **74**, 3273 (1995)

[13] W.H. Meiklejohn and C.P. Bean, Phys. Rev. **102**, 1413 (1956)

[14] W.H. Meiklejohn and C.P. Bean, Phys. Rev. **105**, 904 (1957)

[15] J. Nogués and I.K. Schuller, J. Magn. Magn. Mater. **192**, 203 (1999)

[16] T. Lin, D. Mauri, N. Staurd, C. Huang, JK. Howard, and G.L. Gorman, Appl. Phys. Lett. **65**, 1183 (1994)

[17] S. Mao, S. Gangopadhay, N. Amin, and E. Murdock, Appl. Phys. Lett. **69**, 3593 (1996)

[18] H.S. Jung, W.D. Doyle, and S. Matsunuma, J. Appl. Phys. **93**, 6462 (2003)

[19] Y. Lamy and B. Viala, J. Appl. Phys. **97**, 10F910 (2005)

[20] Y. Lamy, B. Viala, and I.L. Prejbeanu, IEEE Trans. Magn. **41**, 3517 (2005)

[21] Y. Lamy and B. Viala, IEEE Trans. Magn. **42**, 3332 (2006)

[22] Y. Lamy, PhD Thesis, Université de Limoges (2006)

[23] J. Dooley and M. De Graef, Ultramicroscopy **67**, 113 (1997)

[24] A. C. Daykin, J.P. Jakubovics and A. K. Petford-Long, J. Appl. Phys. **82**, 2447 (1997)

[25] W. J. S. Blackburn, G. H. Cutis and R. P. Ferrier, J. Phys. E **2**, 570 (1969)

[26] M. R. Fitzsimmons, P. Yashar, C. Leighton, I. K. Schuller, J. Nogues, C. F. Majkrzak, and J. A. Dura, Phys. Rev. Lett. **84**, 3986 (2000)

[27] C. Phatak, J.A. Bain, J.G. Zhu and M. De Graef, Microscopy and Microanalysis **14**, 832-833 (2008)

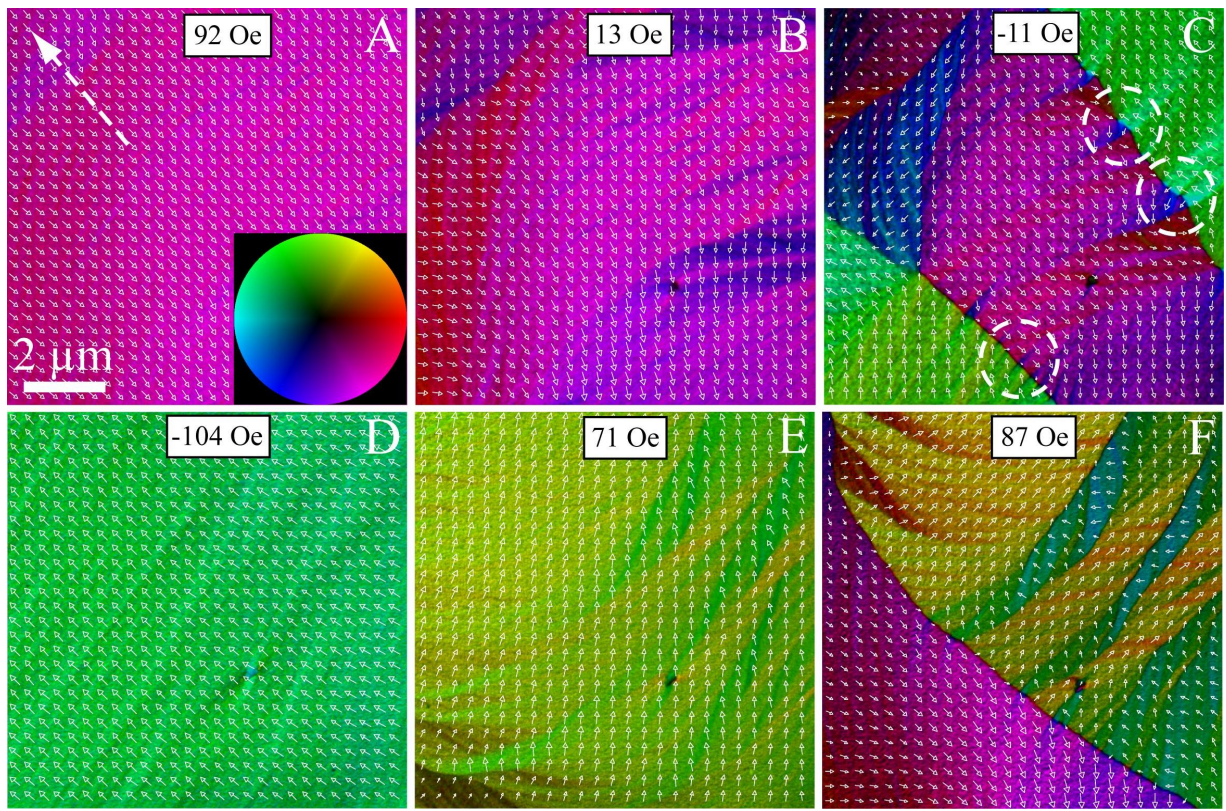

FIG. 1. 6 magnetic maps obtained by DPC on the $Co_{35}Fe_{65}$(70nm)/NiMn(50nm) bilayer. On the image (A) the unidirectional anisotropy direction is represented by the dashed white arrow. The inserted color wheel indicates the direction of magnetization by the color and the intensity corresponds to the strength of the magnetic signal. On the image (C) some cross-tie walls are surrounded.

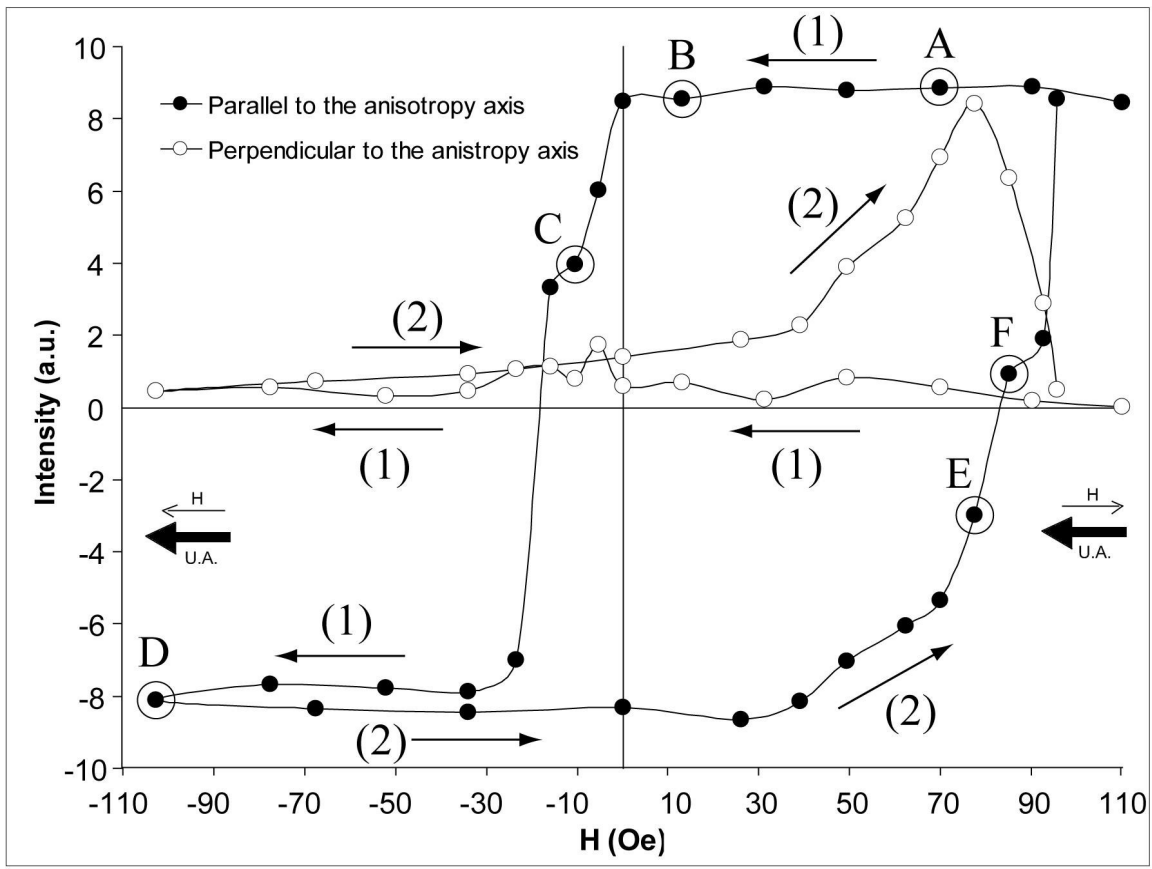

FIG. 2. Hysteresis loops calculated from whole set of magnetic maps for different *in-situ* applied fields. The solid lines are guides to the eye. Magnetic paths (1 and 2) described in the text and locations of magnetic maps on Figure 1 are indicated. Applied filed with respect to the Uniaxial Anisotropy (UA) axis is also given.